\documentclass[preprint2]{aastex6}

\usepackage{graphics,graphicx,subfigure}

\begin{document}


\title{submillimeter polarization observation of the protoplanetary disk around HD 142527}


\author{Akimasa Kataoka\altaffilmark{1,2}, Takashi Tsukagoshi\altaffilmark{3}, Munetake Momose\altaffilmark{3}, Hiroshi Nagai\altaffilmark{2}, Takayuki Muto\altaffilmark{4}, Cornelis P. Dullemond\altaffilmark{1}, Adriana Pohl \altaffilmark{1,5}, Misato Fukagawa\altaffilmark{6}, Hiroshi Shibai\altaffilmark{7}, Tomoyuki Hanawa\altaffilmark{8}, \and Koji Murakawa\altaffilmark{9}}
\altaffiltext{1}{Zentrum f\"ur Astronomie der Universit\"at Heidelberg, Institut f\"ur Theoretische Astrophysik, Albert-Ueberle-Str. 2, D-69120 Heidelberg, Germany \email{kataoka@uni-heidelberg.de}}
\altaffiltext{2}{National Astronomical Observatory of Japan, Mitaka, Tokyo 181-8588, Japan}
\altaffiltext{3}{College of Science, Ibaraki University, 2-1-1 Bunkyo, Mito, Ibaraki 310-8512, Japan}
\altaffiltext{4}{Division of Liberal Arts, Kogakuin University, 1-24-2 Nishi-Shinjuku, Shinjuku-ku, Tokyo 163-8677, Japan}
\altaffiltext{5}{Max Planck Institute for Astronomy, K\"onigstuhl 17, D-69117 Heidelberg, Germany}
\altaffiltext{6}{Division of Particle and Astrophysical Science, Graduate School of Science, Nagoya University, Furo-cho, Chikusa-ku, Nagoya, Aichi 464-8602, Japan}
\altaffiltext{7}{Graduate School of Science, Osaka University, 1-1 Machikaneyama, Toyonaka, Osaka 560-0043, Japan}
\altaffiltext{8}{Center for Frontier Science, Chiba University, 1-33 Yayoi-cho, Inage, Chiba 263-8522, Japan}
\altaffiltext{9}{College of General Education, Osaka Sangyo University, 3-1-1, Nakagaito, Daito, Osaka 574-8530, Japan}




\begin{abstract}
We present the polarization observations toward the circumstellar disk around HD 142527 by using Atacama Large Millimeter/submillimeter Array (ALMA) at the frequency of 343 GHz.
The beam size is $0 \farcs51 \times 0\farcs44$, which corresponds to the spatial resolution of $\sim$ 71 $\times$ 62 au.
The polarized intensity displays a ring-like structure with a peak located on the east side with a polarization fraction of $P= 3.26 \pm 0.02\%$ , which is different from the peak of the continuum emission from the northeast region.
The polarized intensity is significantly weaker at the peak of the continuum where $P= 0.220 \pm 0.010 \%$.
The polarization vectors are in the radial direction in the main ring of the polarized intensity, while there are two regions outside at the northwest and northeast areas where the vectors are in the azimuthal direction.
If the polarization vectors represent the magnetic field morphology, the polarization vectors indicate the toroidal magnetic field configuration on the main ring and the poloidal fields outside.
On the other hand, the flip of the polarization vectors is predicted by the self-scattering of thermal dust emission due to the change of the direction of thermal radiation flux.
Therefore, we conclude that self-scattering of thermal dust emission plays a major role in producing polarization at millimeter wavelengths in this protoplanetary disk.
Also, this puts a constraint on the maximum grain size to be approximately 150 ${\rm \mu m}$ if we assume compact spherical dust grains.
\end{abstract}

\keywords{polarization --- protoplanetary disks --- stars: individual (HD 142527)}



\section{Introduction} \label{sec:intro}

Polarized emission at (sub)millimeter wavelengths has been observed from star-forming regions \citep[e.g.,][]{Girart06, Rao09, Girart09, Hull13, Hull14}. 
As a direct extension, there have been several observations trying to detect (sub)millimeter-wave polarization of circumstellar disks.
It has been difficult to detect (sub)millimeter polarization in protoplanetary disks with single-dish telescopes because of a lack of spatial resolution \citep[e.g.,][]{Tamura95}.
For cases of young circumstellar disks embedded in their envelopes, by using radio interferometry with high spatial resolution, there are some detections of (sub)millimeter-wave polarization toward IRAS 16293-2422 B \citep{Rao14}, L 1527 \citep{Segura-Cox15}, and HL Tau \citep{Stephens14}.
However, there has been no detection of millimeter-wave polarization signature from circumstellar disks in the later stages such as HD 163296, TW Hya \citep{Hughes09}, GM Aur, DG Tau, and MWC 480 \citep{Hughes13}.

Two possible mechanisms have been proposed for the polarization at millimeter wavelengths.
One is the grain alignment with magnetic fields.
If dust grains are elongated, they could align perpendicular to the magnetic fields \citep{DavisGreenstein51}.
Radiative torque can spin up the dust grains to help the alignment \citep{LazarianHoang07}.
Therefore, the polarization vectors have been used as a tracer of the magnetic fields.
Particularly in protoplanetary disks, the morphology of the magnetic fields is usually assumed to be toroidal \citep[e.g.,][]{Brandenburg95}, and thus the polarization vectors are predicted to be directed radially \citep{ChoLazarian07}.

The other mechanism is the self-scattering of thermal dust emission.
If dust grains grow to sizes as large as the observed wavelengths and if the radiation field at the same wavelengths has an anisotropy, the thermal dust emission can be scattered by the dust grains themselves and make the emission polarized \citep{Kataoka15}.
For example, the polarized emission from HL Tau disk can be explained either by the grain alignment with complex magnetic field morphology \citep{Stephens14} or by the self-scattering \citep{Kataoka16,Yang16a}.
In the case of face-on disks, the polarization vectors of disks are predicted to be in the radial direction at inner radii but in the azimuthal direction at outer radii \citep{Kataoka15}.

In this Letter, we report the first detection of millimeter-wave polarization from a late-stage protoplanetary disk.
The target is HD 142527, which is a Herbig Ae star \citep{Waelkens96}.
The associated circumstellar disk has a wide gap with spiral patterns in infrared observations \citep{Fukagawa06, Avenhaus14} and a lopsided continuum structure with submillimeter-wave observations \citep{Casassus13, Fukagawa13}, which possibly show the trapping of dust by a vortex \citep[e.g.,][]{Birnstiel13, ZhuStone14}.

\begin{figure*}[ht!]
\figurenum{1}
\plottwo
{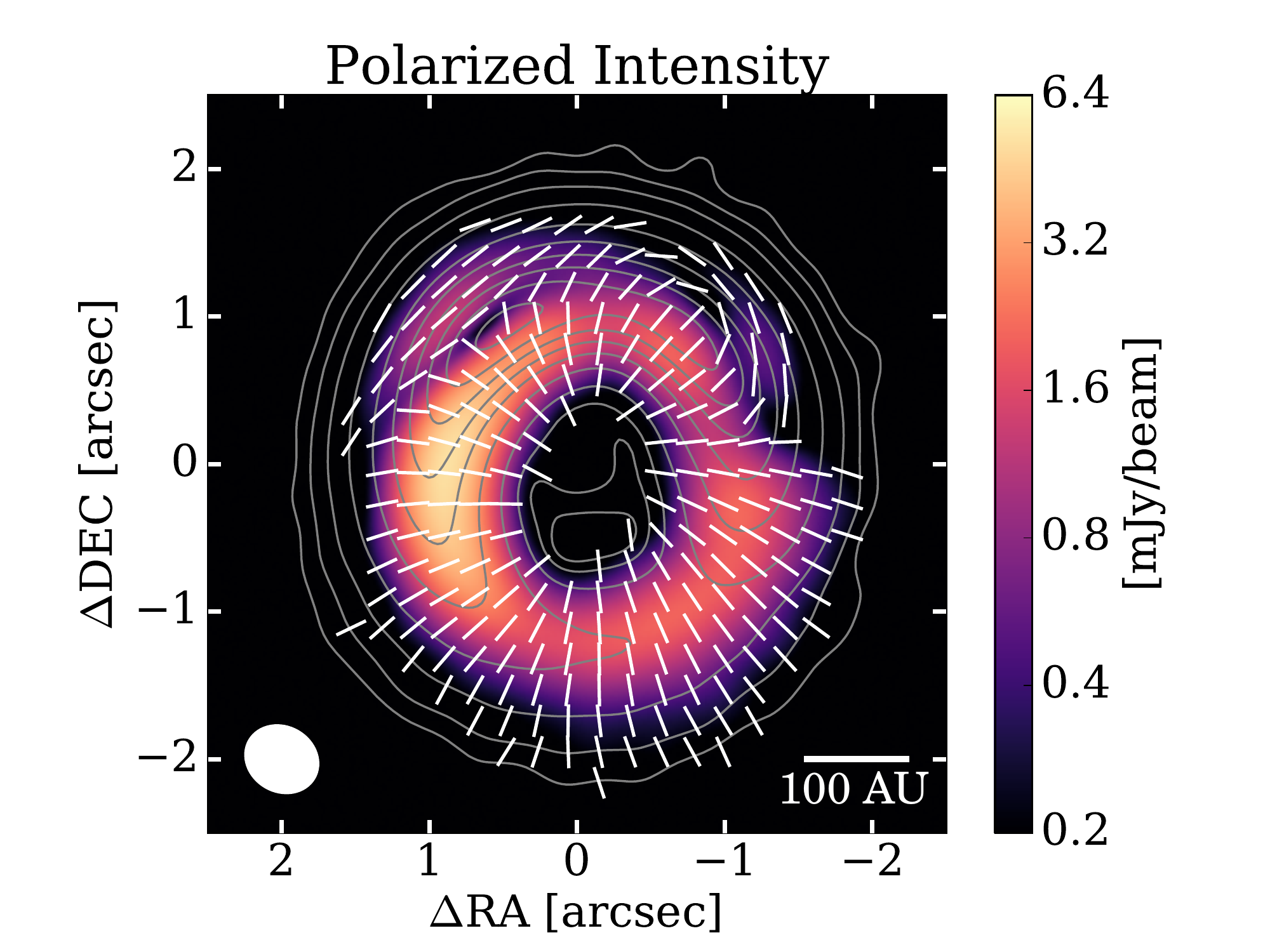}
{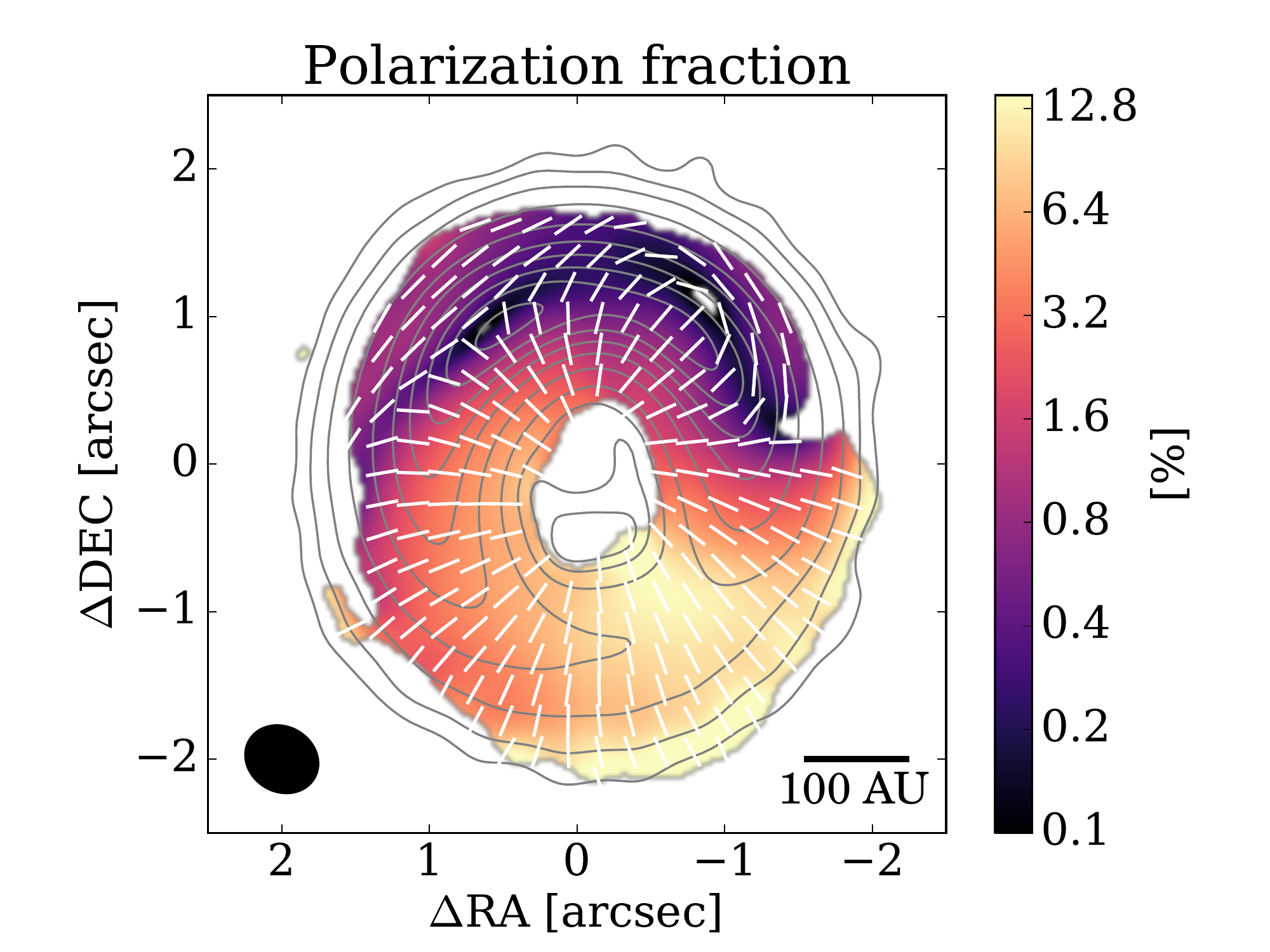}
\caption{
In the left panel, the colorscale represents the polarized intensity in units of ${\rm~mJy~beam^{-1}}$ with a log scale, the gray contours show the continuum emission, and the white vectors show the polarization vectors.
Note that the lengths of the polarization vectors are set to be the same.
The levels of the contours are $(3, 10, 30, 100, 300, 600, 900, 1200, 1500,1800) \times \sigma_{\rm I} (=185 {\rm~\mu Jy~beam^{-1}})$ for Stokes $I$. 
Polarization vectors are plotted where the polarized intensity is larger than $3\sigma_{\rm PI}=0.128 {\rm~mJy~beam^{-1}}$.
In the right panel, the colorscale displays the polarization fraction overlaid with the polarization vectors.
The gray contours display the continuum emission with the same levels of the left panel.
The colorscale is only shown with the same threshold of the polarization vectors in the left panel.
\label{fig:PI}}
\end{figure*}

\begin{figure*}[ht!]
\figurenum{2}
\centering
\subfigure{\includegraphics[width=80mm]{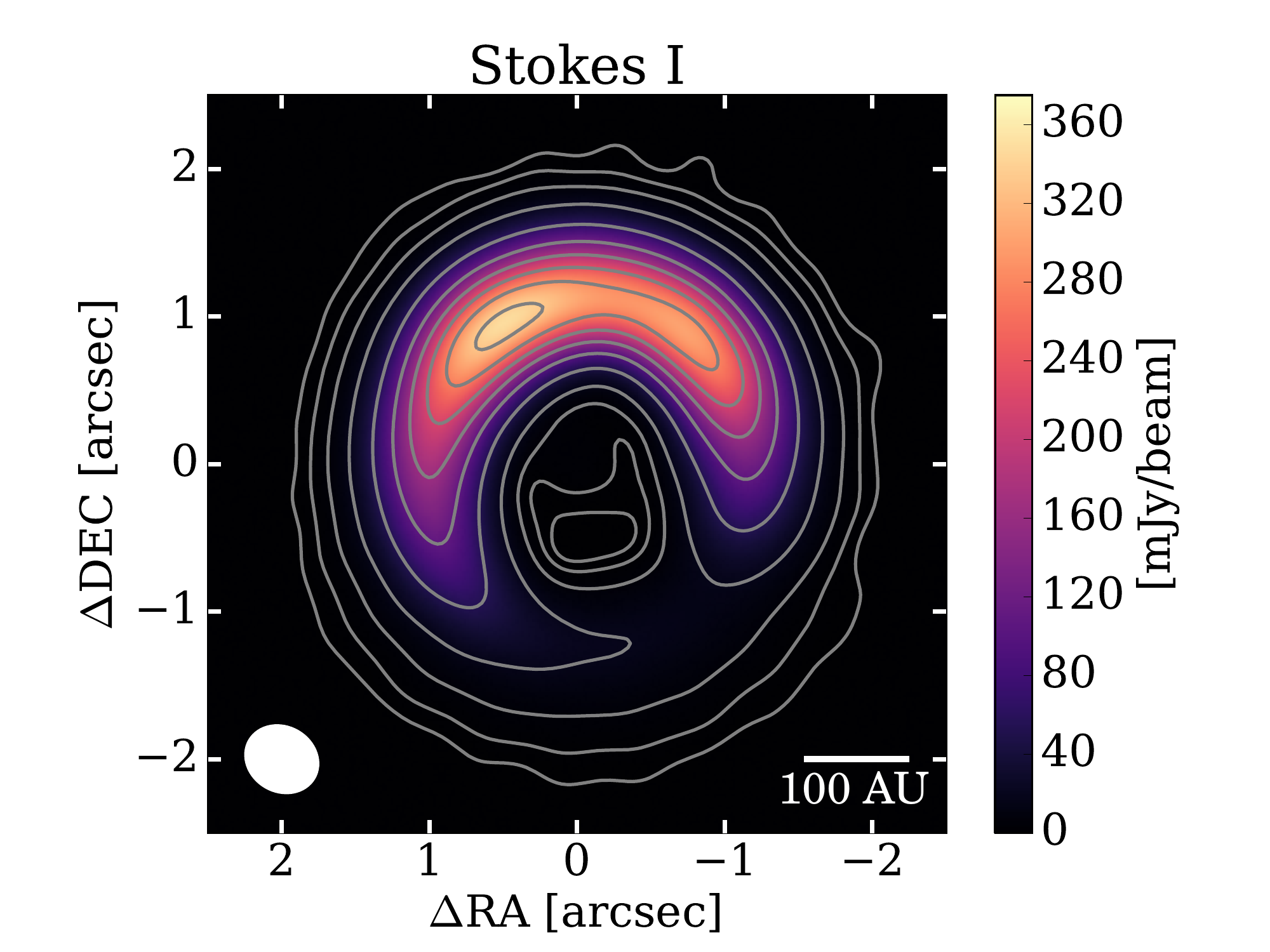}}
\subfigure{\includegraphics[width=80mm]{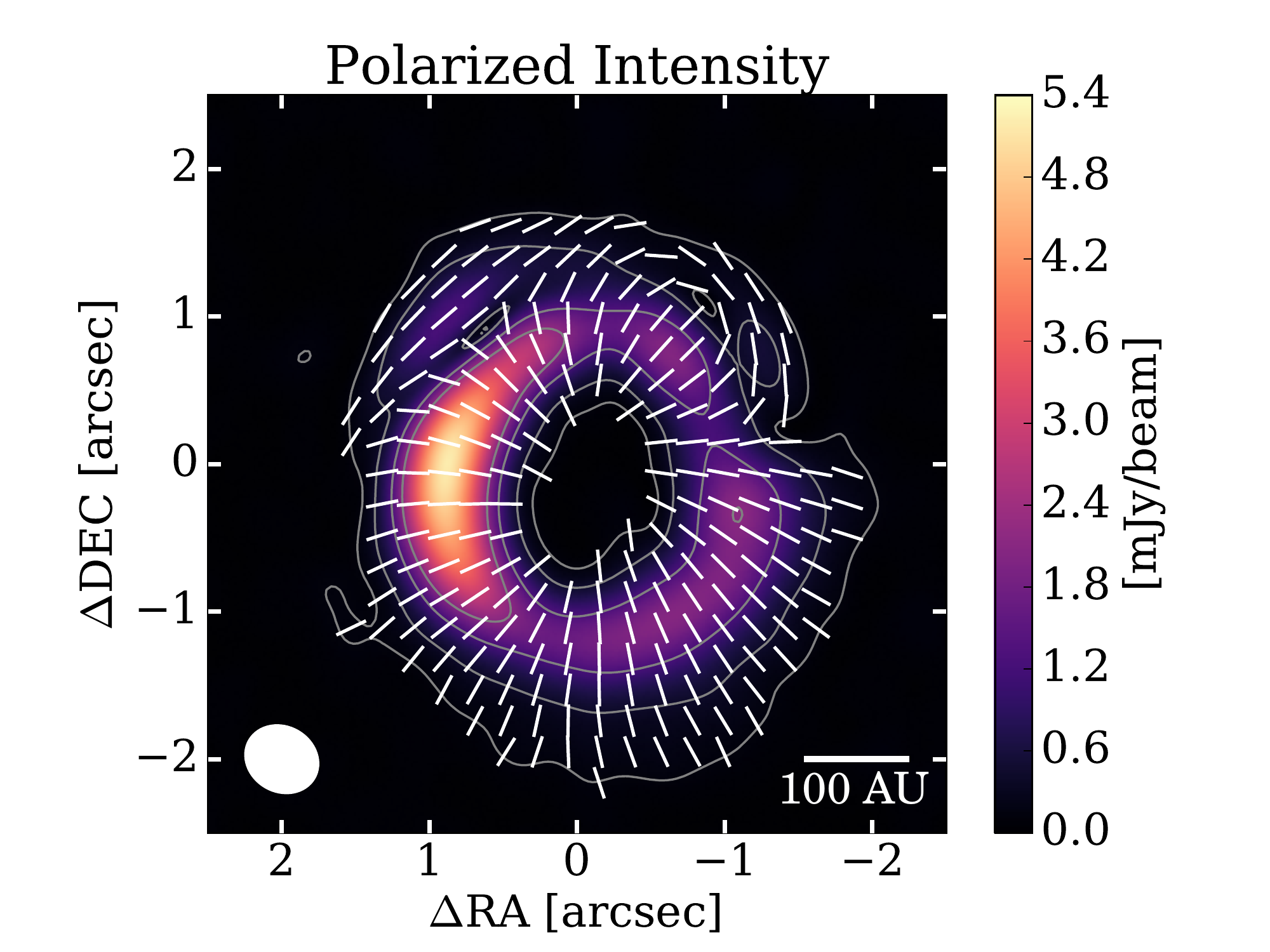}}\\
\subfigure{\includegraphics[width=80mm]{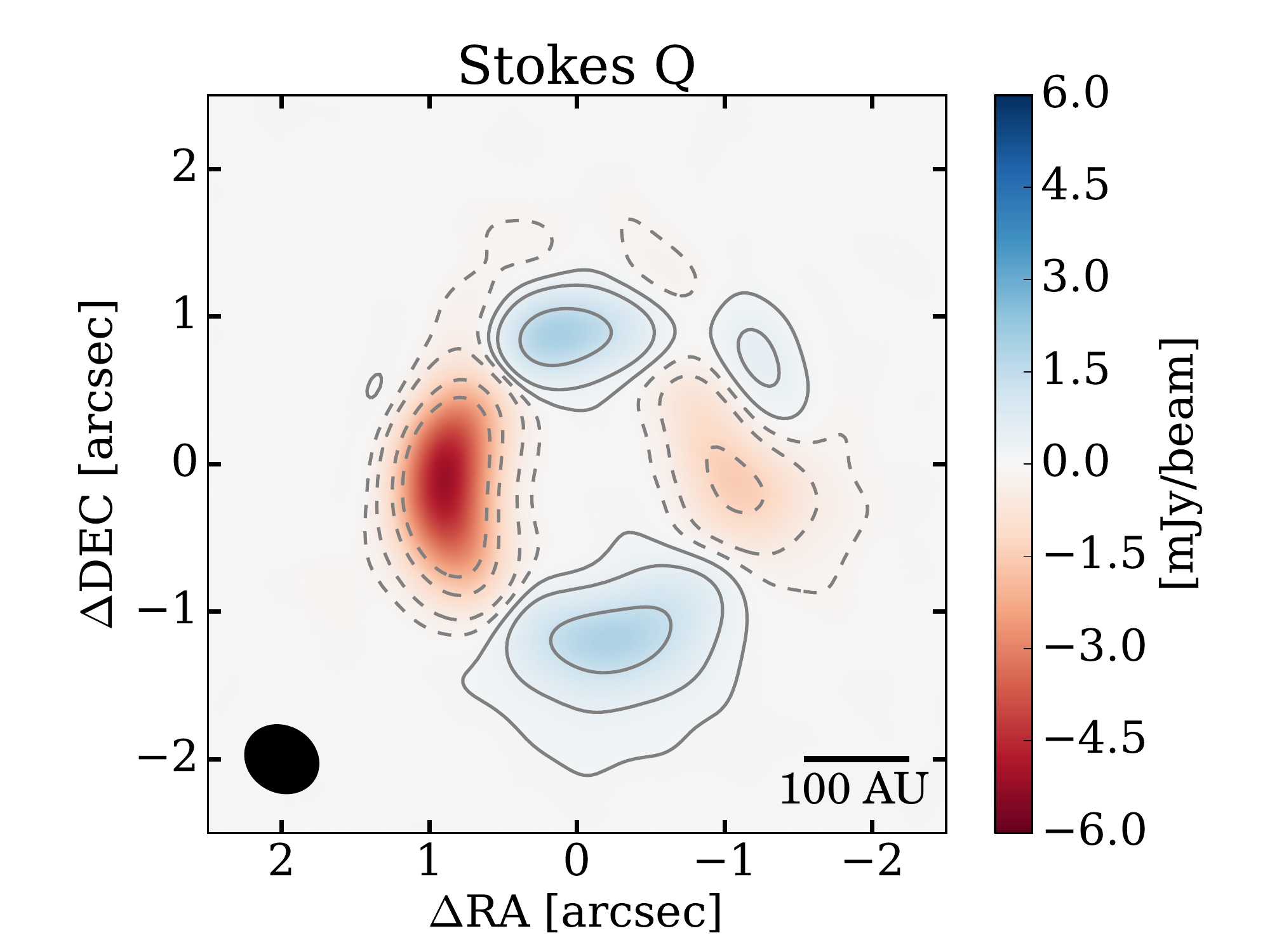}}
\subfigure{\includegraphics[width=80mm]{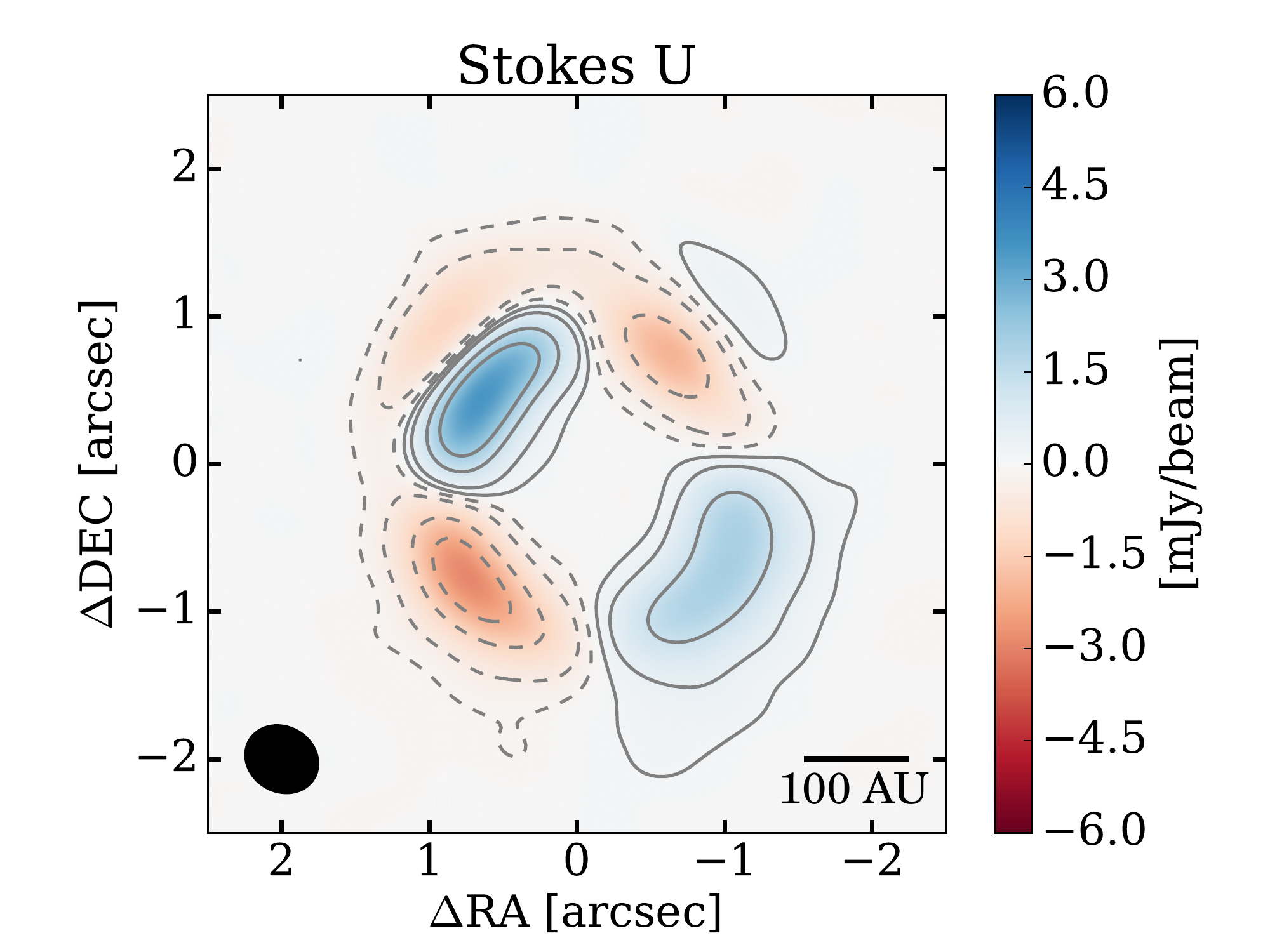}}
\vspace{10pt}
\caption{
Four maps display the intensity of Stokes $I$, polarized intensity, Stokes $Q$, and Stokes $U$ maps in a linear scale.
Levels of contours are $(3, 10, 30, 100, 300, 600, 900, 1200, 1500,1800) \times \sigma_{\rm I} (= 185 {\rm~\mu Jy~beam^{-1}})$ for Stokes $I$, $(3, 10, 30, 50, 100)\times \sigma_{\rm PI} (= 42.8 {\rm~\mu Jy~beam^{-1}})$ for the polarized intensity, and $(-50, -30,-10,-3,3,10,30) \times \sigma_{\rm PI}$ for Stokes $Q$ and $(-50, -30,-10,-3,3,10,30, 50) \times \sigma_{\rm PI}$ for Stokes $U$.
The solid contours in the Stokes $Q$ and $U$ maps represent plus components, and the dashed contours represent minus components.
\label{fig:IQUV}}
\end{figure*}

\section{Observations}
HD142527 was observed by Atacama Large Millimeter / submillimeter Array (ALMA) on 2016 March 11, during its Cycle 3 operation.
The antenna configuration was C36-2/3 and 38 antennas were operated during the observation period.
The correlator processed four spectral windows centered at $336.5, 338.4, 348.5,$ and $350.5$ GHz with a bandwidth of 1.875 GHz each.
The bandpass and the complex gain were calibrated by observations of J1427-4206 and J1604-4441, respectively, and the polarization calibration was performed by observations of J1512-0905.
The raw data were reduced by EA-ARC staff.
A detailed description of the data reduction is given in \citealt{Nagai16}.

We further perform the iterative CLEAN deconvolution imaging by using CASA with self-calibration to improve the image quality.
We employ the Briggs weighting with a robust parameter of $0.5$ and the multiscale option with scale parameters of 0, 0.3, and 0.9 arcsec.
The beam size of the final product is $0 \arcsec.51 \times 0 \arcsec.44$, corresponding to $\sim71\times 62$ au at the distance to the target.
The rms for each Stokes component is summarized in Table \ref{tab:obs}.
The rms for the Stokes $V$ image is 25.2 $\mu$Jy, close to the thermal noise level of the integration time.
Since the Stokes $I$, $Q$, and $U$ images are dynamic range limited, the high rms on Stokes $I$, $Q$, and $U$ images is not due to the lack of integration time nor the polarization calibration error.

\begin{deluxetable*}{c|cccc}
\tablecaption{Observational Data of Each Stokes Component \label{tab:obs}
}
\tablehead{
& \colhead{Total Flux (mJy)}& \colhead{Max, Min (mJy beam$^{-1}$)}& \colhead{rms ($\mu$Jy beam$^{-1}$)}  
}
\startdata
Stokes $I$ & 847 &  340 & 185 \\
Stokes $Q$ & ... & 1.99, -5.13 & 29.1 \\
Stokes $U$ & ... & 3.52, -2.88 & 29.5  \\
Polarized Intensity & 1.75 & 5.22 & 42.8\\
\enddata
\tablecomments{ The Stokes $I$ and the polarized intensity show the maximum flux while the Stokes $Q$ and $U$ show both maximum and minimum values in the middle column.}
\end{deluxetable*}

\section{Results}

The total and peak fluxes of each Stokes component are also summarized in Table \ref{tab:obs}.
Here, the maximum flux is shown for the Stokes $I$ while both maximum and minimum fluxes are shown for the Stokes $Q$ and $U$.
Spatially integrated Stokes $Q$ and $U$ components give the total polarized intensity $PI=\left( (\sum Q)^2 + (\sum U)^2 \right)^{0.5}$ to be $1.75$ mJy with an rms of $\sigma_{\rm PI}=42.8 {\rm~\mu Jy~beam^{-1}}$.
Dividing the total polarized intensity by the total intensity, we derive the total polarization fraction to be $0.207$ \%.
This low fraction of polarization is unlikely to be detected if the disk is spatially unresolved.
Therefore, the high spatial resolution of this observation is essential to detect the polarization.

The left panel of Figure \ref{fig:PI} shows the polarized intensity by colorscale overlaid with the solid contour of the continuum and the polarization vectors.
The continuum emission well reproduces the previously observed lopsided structure \citep[e.g.,][]{Casassus13,Fukagawa13}.
The polarized intensity shows a ring-like distribution with azimuthal asymmetry, for which the substructure is different from the continuum, and also shows two regions with azimuthal polarization.
Although the continuum emission has a peak at the northeast region, the polarized intensity has a peak on the east side.
The peak emission of the polarized emission is $5.22$ ${\rm mJy~beam^{-1}}$.
The ring of the polarized intensity is located slightly inside of the ring center of the continuum.
The maps of each Stokes component are shown in Fig. \ref{fig:IQUV}.

The polarization vectors on the main ring of the polarized emission have a radial direction everywhere.
However, the polarization vectors are rotated by $90^\circ$ in two regions.
The northeast region is $\sim 1\farcs3 $ from the star toward the position angle of $\sim 42^\circ$  while the northwest one is $\sim 1\farcs4$ from the star toward the position angle of $\sim -60^\circ$. 
The two regions are clearly seen in Fig. \ref{fig:IQUV} as a flip of the sign of Stokes $Q$ or $U$ from inside to outside.

Furthermore, the polarized intensity is relatively bright in the southwest direction, where Stokes I is the faintest.
This causes high polarization fraction in the south region.
The right panel of Fig. \ref{fig:PI} shows the polarization fraction overlaid with the polarization vectors with the continuum as solid contours.
The polarization fraction is $3.26 \pm0.02$ \% at the peak of the polarized intensity and as low as $0.220 \pm 0.010$ \% at the peak of the continuum.
The polarization fraction has a peak at the southwest region with a fraction of $13.9 \pm 0.6$ \%, which corresponds to around the local minimum of the intensity of the main ring.

Here, we note that the polarization maps at low signal-to-noise ratio regions could be affected by the positive polarization bias \citep{Vaillancourt06}.
The discussion in this Letter is concerning at the regions where the detection is larger than 3 $\sigma_{\rm PI}$ and thus the positive polarization bias does not affect the results.

\section{Discussions}
We have detected spatially resolved polarized continuum emission from the disk around HD 142527.
There are three distinct observational signatures - (1) difference of the locations of the brightest emission between Stokes $I$ and PI, (2) 90$^\circ$ flip of the polarization vectors in the northeast and northwest region, and (3) the high fraction of polarization ($13.9 \pm 0.6$\%) in the southwest region.
Two possible mechanisms to produce polarized emission in protoplanetary disks are suggested to date: grain alignment by the magnetic field or dust self-scattering.
In this section, we qualitatively discuss which mechanism is more likely to take place in the disk around HD 142527.  

\subsection{Grain alignment}

Here, we discuss the possibility that the polarization is due to the grain alignment with magnetic fields.
The magnetic field direction is rotated by $90^\circ$ from the polarization vectors in the thermal emission regime.
Therefore, the morphology of the main polarization ring indicates the presence of toroidal magnetic fields, which is consistent with the common understanding of the magnetic field in disks \citep[e.g.,][]{Brandenburg95, Stone96}.

The difference between the peak position of the polarized intensity and the peak of the continuum could be explained with the depolarization due to the high optical depth at the peak of the continuum \citep[e.g.,][]{Alves14}.
We will also discuss the effects of the optical depth in the next section.
Another possibility is the difference in the grain size and its effects on the alignment efficiency; the radiative torque efficiency decreases with decreasing grain size if the grain size is smaller than the wavelength.
Therefore, if the peak emission of Stokes I is mainly coming from grains smaller than the wavelengths, it would decrease the alignment efficiency and thus the polarization fraction \citep[e.g.,][]{ChoLazarian07, LazarianHoang07}.

The azimuthal direction of the polarization vectors as shown in the outer regions indicates the poloidal magnetic field configuration, while the radial direction on the inner ring indicates the toroidal magnetic field.
There has been no mechanism locally to rotate the direction of the magnetic field by $90^\circ$.
Therefore, at least at that position, the mechanism should be different from the grain alignment.

The polarization fraction in the south region is as high as $13.9 \pm 0.7$ \%, which is higher than the predicted value \citep[e.g.,][]{ChoLazarian07}.
The high fraction of polarization means that the alignment efficiency maybe higher than expected or the long-to-short axis ratio of elongated dust grains is larger than the assumed value \citep{ChoLazarian07}.
Alternatively, the high polarization fraction observed in the southwest region could be due to interferometric filtering effects, where the Stokes $I$ and the Stokes $Q$, $U$ maps are resolved out differently.

Here, we also note that the high polarization fraction observed in the southwest region could be due to interferometric effects where the Stokes $I$ and the Stokes $Q$, $U$ maps are resolved out.

\begin{figure*}[ht!]
\figurenum{3}
\centering
\subfigure{\includegraphics[width=80mm]{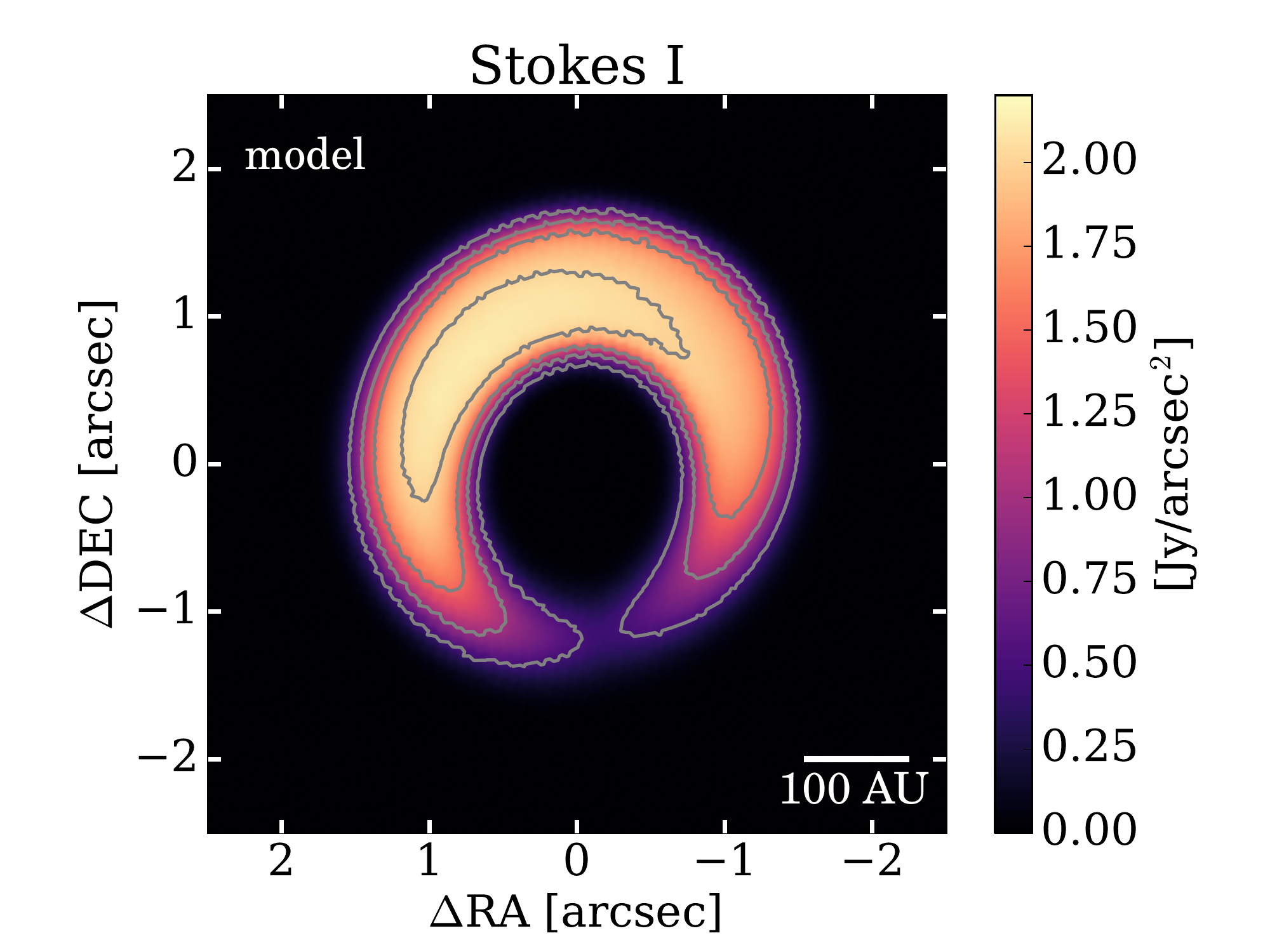}}
\subfigure{\includegraphics[width=80mm]{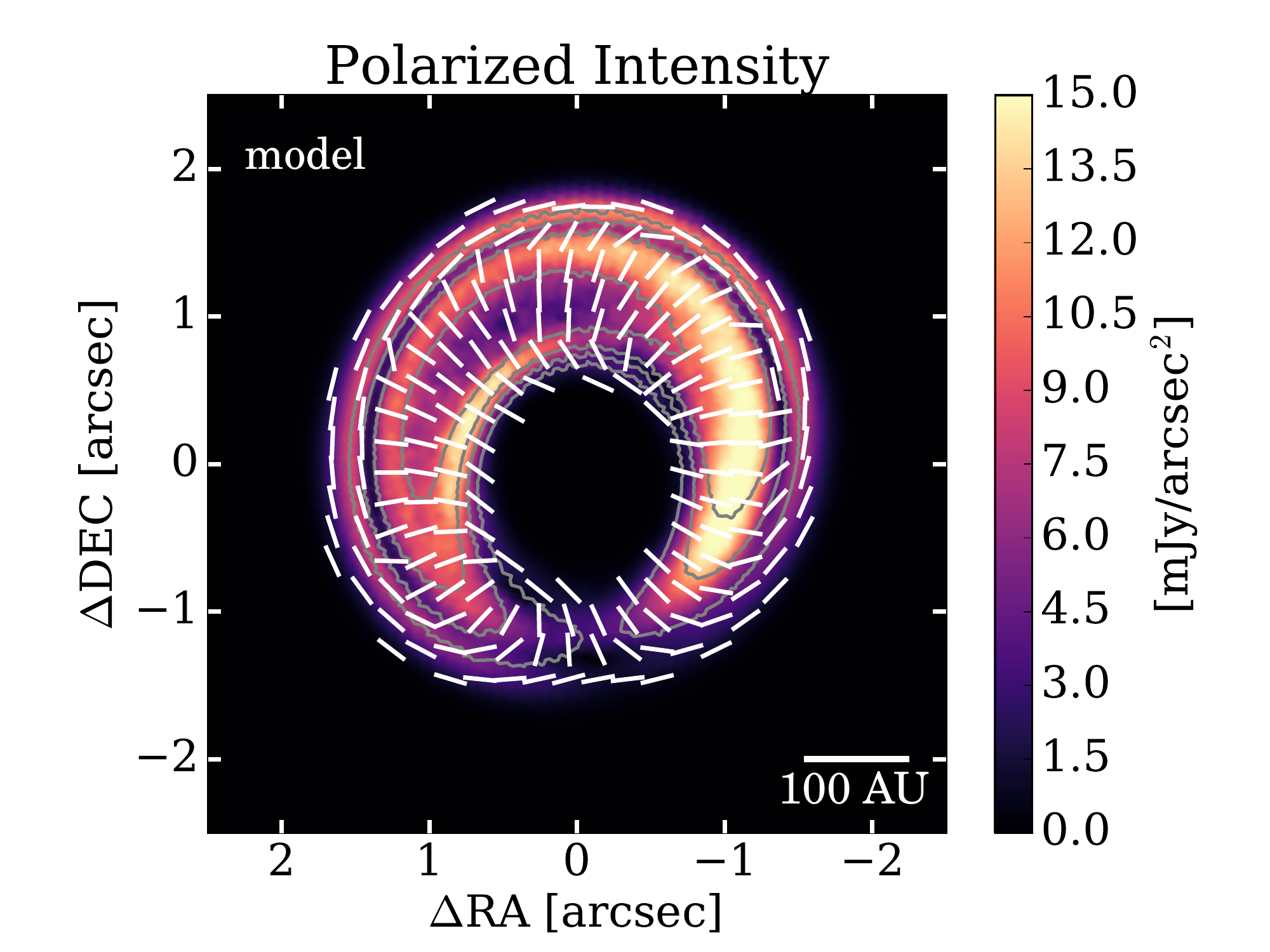}}
\subfigure{\includegraphics[width=80mm]{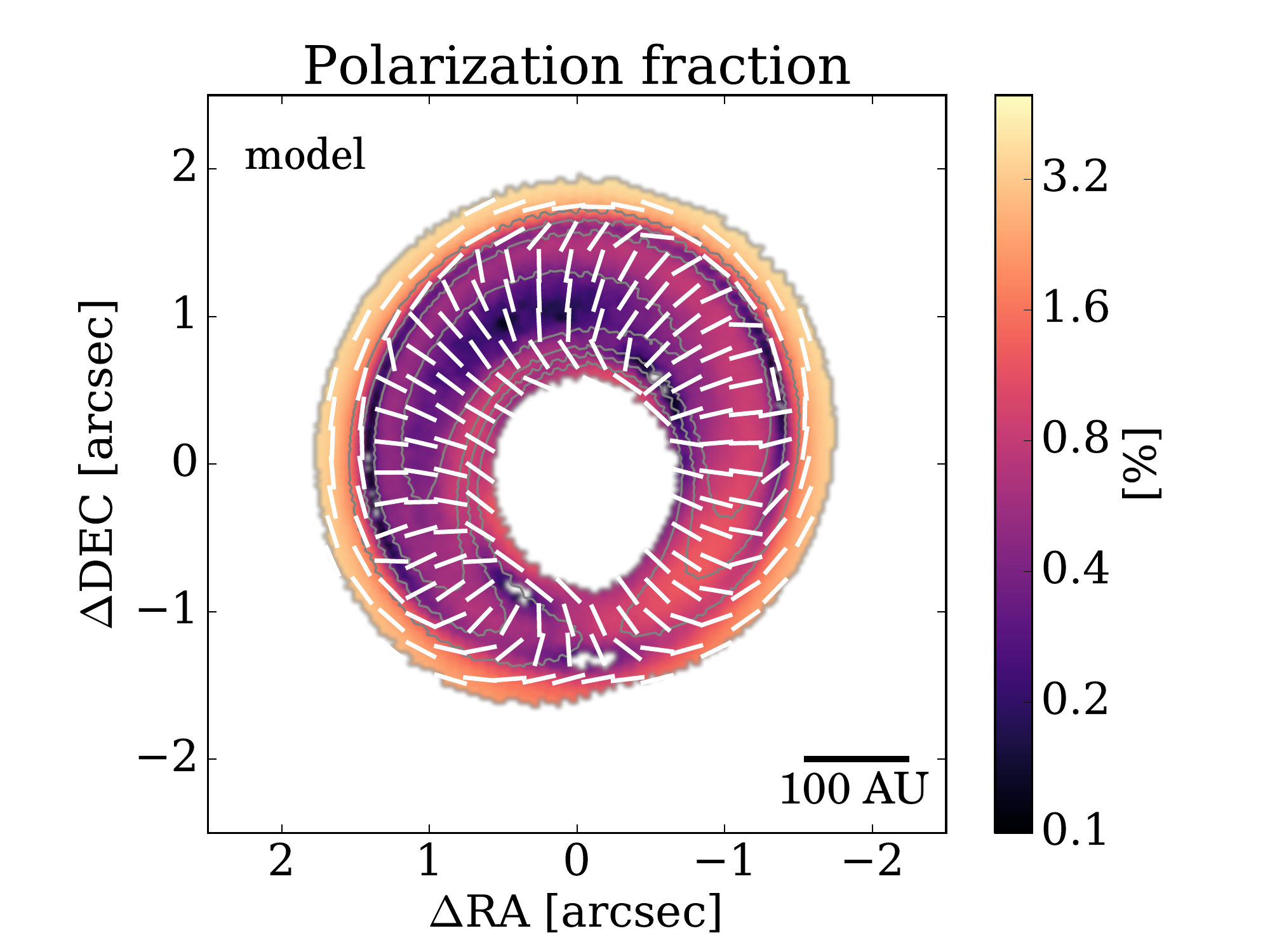}}
\caption{ 
Colorscale of each figure displays the intensity, the polarized intensity, and the polarization fraction of the model calculations.
The levels of the solid contours are $(5, 10, 15, 20)\times \sigma_{\rm I, sim}  (= 0.1 {\rm~ Jy~arcsec^{-2}) }$ of the intensity.
}
\label{fig:model}
\end{figure*}

\subsection{Self-scattering}

\subsubsection{Model Prediction}
Another possible explanation for the millimeter-wave polarization is the self-scattering of the thermal dust emission at the observed wavelengths \citep{Kataoka15}.
We model the intensity with a simple model and perform radiative transfer calculations with RADMC-3D \footnote{RADMC-3D is an open code of radiative transfer calculations. The code is available online: http://www.ita.uni-heidelberg.de/\~{}dullemond/software/radmc-3d/} to see the model prediction of the self-scattering.
We assume that the dust grains have a power-law size distribution with a power of $-3.5$ and the maximum grain size $150 {\rm~\mu m}$, which is the most efficient grain size to scatter the thermal emission and make it polarized at the observed wavelength of $\lambda=0.87$ mm.
The density distribution is based on a previous modeling of the continuum at the same wavelength \citep{Muto15}.
The density distribution of dust grains $\Sigma_{\rm d}$ is taken to be a Gaussian distribution in radial and azimuthal direction as
\begin{eqnarray}
\Sigma_{\rm d} = \left[\Sigma_{\rm 0,min} + (\Sigma_{\rm 0,max}-\Sigma_{\rm 0,min})\exp\left(-\frac{(\theta-\theta_d)^2}{2\phi_d^2}\right)\right]\nonumber\\ 
\times \exp\left(-\frac{(r-r_c)^2}{2r_w^2}\right),
\end{eqnarray}
where $r$ and $\theta$ are the radial and azimuthal coordinates $\Sigma_{\rm 0,max}$ and $\Sigma_{\rm 0,min}$ are the maximum and minimum surface density at the center of the ring, $\theta_d$ and $\phi_d$ represent the center and width of the vortex in the azimuthal direction, and $r_c$ and $r_w$ represent the center and width of the vortex in the radial direction.
The values are taken to be $\Sigma_{\rm 0,max}=0.6{\rm~g~cm^{-2}}$, $\Sigma_{\rm 0,min}=0.008{\rm~g~cm^{-2}}$, $\theta_d=30^\circ$, $\phi_d = 60^\circ$, $r_c = 170$ au, and $r_w=27$ au.
The dust temperature is set to 36 K.
In the calculations of RADMC-3D, we set the inclination of the disk to be $27^\circ$ \citep{Fukagawa13} and the position angle to be $71^\circ$.
Here, we do not aim to model the intensity and the polarization perfectly, but to investigate the morphology and the strength of the polarization with a simple model of an inclined lopsided-disk to understand that feature can be explained with the simple model.

Here, we summarize the model prediction by the self-scattering model.
Figure \ref{fig:model} shows the intensity, the polarized intensity, and the polarization fraction of the model.
The polarized intensity shows the main ring inside with radial polarization vectors and the ring outside with azimuthal polarization vectors.
The polarization fraction is up to $\sim 2\%$ in the inner ring while $\sim 5\%$ in the outer ring.
Also, the polarized intensity is weaker at the peak of the continuum than in other regions because the dust thermal emission is optically thick at the peak of the continuum.
The reason is as follows.
With the polarization observations, we are looking at the emission scattered by dust grains.
Let us consider a dust grain as a scatterer.
The polarization fraction is determined by the superposition of the scattered emission of all incoming fluxes from any direction.
If the thermal dust emission is optically thin, the anisotropy of the radiation field causes difference of the incoming fluxes depending on its direction, that is why we can see the polarization as a residual.
If the thermal dust emission is marginally optically thick, however, there is no difference in incoming fluxes depending on its direction, which causes no residual polarization because they are canceled.
This is why we see less polarization fraction at the optically thick regions.
These features have been predicted by \citealt{Kataoka15} as the face-on ring model.
In the case of HD 142527, the emission from the peak of the continuum is shown to be optically thick at the observed wavelengths \citep{Muto15}.
Therefore, the depolarization due to the optical thickness is expected in the disk.

\subsubsection{Comparison with the Observation}

The flip of the polarization vectors of the observations is naturally explained with the self-scattering model.
This flip is strong support for the self-scattering hypothesis.
The peak difference between the PI and continuum can also be naturally explained by the effects of optical depth.

However, the simple model of the self-scattering with the constant maximum grain size does not explain the observed polarized intensity distribution qualitatively.
The observation shows that the polarized intensity is the brightest at the position angle of $\sim 90^\circ$ while it is faint at the position angle of $\sim -45^\circ$.
However, the model does not show the peak at the position angle of $\sim 90^\circ$.
Also, the high polarization fraction of $13.9 \pm 0.6$ \% at the south region has not been expected from the self-scattering model.
The expected polarization fraction is a few percent for the main ring as shown in Fig. \ref{fig:model} (see also \citealt{Kataoka15}).

\subsubsection{Possible Improvements of the Model}

We discuss further modeling in future publications to explain the observation.
Including the spatial grain size distribution is essential to explain the peak difference between the intensity and the polarized intensity.
If the maximum grain size is smaller or larger than $\sim 150 {~\rm \mu m}$, the polarization fraction decreases.
The asymmetry of the polarized intensity between the leading and the trailing sides of the vortex could be correlated with the difference of the trapping size ahead of the vortex \citep{ZhuBaruteau16, BaruteauZhu16}.
Also, temperature distribution could also reduce or enhance the polarization fraction by changing the radiation field.
Note that the temperature at the northwest region may be higher than other areas \citep[e.g.,][]{Casassus15}.
For explaining the high fraction of polarization, unresolved substructure such as a narrower ring should be added, which enhances the anisotropies of the radiation field and thus increases the polarization fraction.

\subsubsection{Indication on Dust Coagulation and Dynamics}

In this Letter, we have shown that the millimeter-wave polarization is explained with grains with the maximum size of $150 {~\mu m}$ because the polarization is effective when the maximum grain size is $\sim \lambda/2\pi $.
The available size range to explain the observations is roughly $\sim 40-300 {\rm~\mu m}$ if we assume the compact spherical grains \citep{Kataoka15}.
This relatively small grain size is not consistent with the previous understanding of the dust trapping that requires larger grains of millimeter or centimeter. 

The questions regarding the understanding of the continuum emission were how to trap the dust grains both radially and azimuthally.
For the effective trapping of dust grains both radially and azimuthally, the grains should be partly decoupled from the gas, which suggests the grain size to be around millimeter to centimeter \citep[e.g.,][]{Pinilla12a, Birnstiel13}.
This theoretical explanation is supported by the observations at millimeter and centimeter wavelengths.
The observations of the protoplanetary disk around HD 142527 by ALMA and ATCA exhibit the low spectral index and the spectral index is lower at the center of the vortex, which suggests that the grain size is around millimeter to centimeter and larger grains are more efficiently trapped \citep{Casassus15}.

Furthermore, if dust grains are radially trapped by a pressure enhancement by a planet, the polarization rings should be radially narrower.
The grain size is changing in radial direction, and the polarization should be detected only where the maximum grain size is around $150 {\rm~\mu m}$, which are two rings radially separated as surrounding the dust trap \citep{Pohl16}.
Therefore, we are now at the stage to reconsider the interpretation of the emission of dust grains.

One possible idea to solve the problem that the spectral index and the polarization indicate the different grain sizes is to include the fluffiness of dust aggregates.
Dust grains coagulate to form fluffy dust aggregates with a filling factor of $10^{-4}$ \citep[e.g.,][]{Kataoka13b}, and they can also explain the absorption opacity but with large scattering opacity \citep{Kataoka14}.
If the dust grains are fluffy, then the scattering efficiency is expected to be much higher than compact grains \citep{Min16,Tazaki16}, and therefore it could explain the emission from everywhere in the disk.
In this way, the millimeter-wave polarization due to the self-scattering raises new insights into our understanding of the grain growth.

\section{Conclusions}

We have observed the circumstellar disk around HD 142527 with the full polarization mode of ALMA.
We have detected the linear polarization of the continuum at the wavelengths of 0.87 mm.
This observation is the first successful polarization detection of a not-embedded protoplanetary disk with ALMA.
The polarized intensity displays a ring structure with a peak on the east side, which is different from the peak position of the continuum.
The polarized intensity has a local minimum at the peak of the continuum.
The polarization vectors are in the radial direction on the main ring, while those are in the azimuthal direction at the outer regions of the disk.
This change of orientation of polarization vectors is predicted in the self-scattering of the thermal emission \citep{Kataoka15} though it is hard to explain from the grain alignment with the magnetic fields because the magnetic fields are not expected to be rotated by 90$^\circ$ in protoplanetary disks.
Therefore, we conclude that the self-scattering plays an important role in millimeter-wave polarization in the protoplanetary disk around HD 142527.
The self-scattered polarization leads us to a new direction of observational tests on the grain growth: we can obtain the information of the absorption opacity from the continuum and the scattering opacity from the polarization, which have never been modeled at the (sub)millimeter wavelength range.

\acknowledgements
We acknowledge the fruitful discussions with Laura Perez, Leonardo Testi, Marco Tazzari, Carlo Manara, Til Birnstiel, and Paola Pinilla.
This work is supported by MEXT KAKENHI No. 23103004 and by JSPS KAKENHI Nos. 15K17606 and 26800106.
This work makes use of the Astropy \citep{AstropyCollaboration13} and {Matplotlib} \citep{Hunter07} software packages.
This Letter makes use of the following ALMA data:
  ADS/JAO.ALMA\#2015.1.00425.S. ALMA is a partnership of ESO (representing its member states), 
  NSF (USA) and NINS (Japan), together with NRC (Canada), NSC and ASIAA (Taiwan), and KASI 
  (Republic of Korea), in cooperation with the Republic of Chile. 
  The Joint ALMA Observatory is operated by ESO, AUI/NRAO and NAOJ.





\end{document}